\documentclass{iauc}
\usepackage{graphicx}
\pubyear{2005}
\volume{199}
\newcommand{\mnras}{MNRAS}
\newcommand{\apj}{ApJ}
\newcommand{\apjl}{ApL}
\newcommand{\araa}{ARAA}
\newcommand{\nat}{Nature}
\newcommand{\aj}{AJ}
\newcommand{\prd}{PhysRevD}
\newcommand{\aap}{A\&A}
\newcommand{\apjs}{ApJS}
\newcommand{\bfx}{{\bf x}}

\newcommand{\bfr}{{\bf r}}

\newcommand{\bfk}{{\bf k}}
\newcommand{\bfv}{{\bf v}}

\newcommand{\bc}{\begin{center}}
\newcommand{\be}{\begin{equation}}
\newcommand{\ee}{\end{equation}}
\newcommand{\ec}{\end{center}}

\newcommand{\boltz}{\hbox{${\rm k}_{\sc\rm B}$}}
\newcommand{\mhydro}{\hbox{$m_{\rm h}$}}
\newcommand{\Mpc}{\rm Mpc}

\newcommand{\cmbfast}{{\sc {CMBFAST }}}

\newcommand{\enzo}{{\sc {ENZO}}}
\newcommand{\gadgetII}{{\sc {Gadget-2}}}

\newcommand{\ltsima}{\mbox{$\; \buildrel < \over \sim \;$}}
\def \simlt{\lower.5ex\hbox{\ltsima}}            % < over ~
\def \gtsima{\mbox{$\; \buildrel > \over \sim \;$}}
\def \simgt{\lower.5ex\hbox{\gtsima}}            % > over ~
\newcommand{\nn}{\nonumber}
\pagerange{1--}
\jname{Probing Galaxies through Quasar Absorption Lines}
\editors{P. R. Williams, C. Shu, and B. M\'{e}nard, eds.}
\title[Simulations of QSO absorbers] %% give here short title %
{Numerical simulations of quasar absorbers}

\author[Theuns]   %% give here short author list %%
{Tom Theuns$^{1,2}$}%

\affiliation{
\noindent $^1$Institute for Computational Cosmology, Department of Physics, University of Durham, UK\break 
\noindent $^2$Department of Physics, University of Antwerp, Belgium}

\begin{document}

\maketitle

\begin{abstract}
The physical state of the intergalactic medium can be probed in great
detail with the intervening absorption systems seen in quasar spectra.
The properties of the Hydrogen absorbers depend on many cosmological
parameters, such as the matter-power spectrum, reionisation history,
ionising background and the nature of the dark matter. The spectra also
contain metal lines, which can be used to constrain the star formation
history and the feedback processes acting in large and small
galaxies. Simulations have been instrumental in investigating to what
extent these parameters can be unambiguously constrained with current
and future data. This paper is meant as an introduction to this
subject, and reviews techniques and methods for simulating the
intergalactic medium.
\end{abstract}

\firstsection % if your document starts with a section,
\section{Introduction}

Quasar spectra contain numerous absorption lines that are produced in
the intervening intergalactic medium (Bahcall \& Salpeter 1965; Gunn \&
Peterson). Most of the lines are transitions of Hydrogen (Lynds 1971;
Sargent et al. 1980), but \lq metal\rq\, transitions, such as for
example due to C, Si, O and many others, are observed as well
(e.g. Cowie et al. 1995).  The Hydrogen lines have column densities
from $N_{\rm H{\sc I}}\approx 10^{11}{\rm cm}^{-2}$ to several times
$10^{20}{\rm cm}^{-2}$, with the number density per unit redshift of
the stronger lines decreasing approximately as a power-law in column
density, see Rauch (1998) for a recent review.\\

The properties of the weaker lines, up to a column density of a few
times $10^{14}{\rm cm}^{-2}$ say at redshift $z=3$, are well reproduced
by numerical simulations of the growth of structure in flat cold-dark
matter dominated cosmological models, in which the intergalactic medium
is ionised by the UV-background produced by galaxies and quasars (Cen
et al. 1994; Zhang, Anninos \& Norman 1995; Miralda-Escud\'e et
al. 1996; Hernquist et al. 1996; Wadsley \& Bond 1996; Zhang et
al. 1997; Theuns et al. 1998; Bryan et al. 1999; Meiksin, Bryan \&
Machacek 2001). In these simulations, the H{\sc I} absorption lines
arise in the mildly over- and under dense filaments of the cosmic web
that forms naturally in this type of cosmology. The evolution of the
number of these lines with redshift, is determined by the decrease in
the mean density of the Universe as function of redshift, and in the
change of the amplitude of the UV-background (Theuns, Leonard \&
Efstathiou 1998; Dav\'e et al. 1999). The impressive agreement between
simulations and data have led to a paradigm shift in the interpretation
of these Lyman-$\alpha$ forest lines, away from the earlier models were
absorption occurred in Lyman-$\alpha$ \lq clouds\rq\, confined by a
variety of mechanisms, to the new model where lines arise in the
slightly denser regions of a continuous, fluctuating intergalactic
medium (IGM), with higher column-density lines arising in higher
density gas, on average. The decomposition of such a continuous field
in terms of (blends of) discrete absorption \lq lines\rq\, is
artificial but the simulations show that the line-properties do depend
on the physical state of the IGM, such as its temperature (Schaye et
al. 1999, 2000b; Ricotti et al. 2000; Bryan \& Machacek 2000; McDonald
et al. 2001), the amplitude of the ionising background (Rauch et
al. 1997; Meiksin \& White 2003; Bolton et al. this volume), the
amplitude and shape of the dark matter power-spectrum (Croft et
al. 1998; Gnedin \& Hamilton 2002; Mandelbaum et al. 2003), on the mass
of the neutrino (Croft et al. 1999) and the dark matter particle
(Narayanan et al. 2000; Viel et al. 2005). This makes the
Lyman-$\alpha$ forest a very valuable probe for a range of physical
properties of the Universe (Hui 1999), given that these mildly
non-linear density fluctuations can be simulated reliably. \\

The fact that the over densities are low also makes it possible to make
simple physical models (Bi, B\"orner \& Chu 1992; Bi \& Davidsen 1997;
Schaye 2001), based on the concept of Jeans smoothing where pressure
gradients cause the gas distribution to differ from the dark matter
distribution on small scales (Theuns, Schaye \& Haehnelt 2000). The
dark matter distribution is approximated as a log-normal field (Coles
\& Jones 1991) or else gas is crafted onto a simulated dark matter
distribution either during the simulation (Gnedin \& Hui 1998) or
afterwards (Viel et al. 2002)\\

Simulated higher column density lines form invariably closer, or even
inside, the high-density dark matter haloes that in the real Universe
are presumably associated with galaxies. The outer parts of the cloud
will shield the inner parts from ionization when the column density is
sufficiently high, $N_{\rm H{\sc I}}\gtsima 10^17\,{\rm cm}^{-2}$. This
self-shielding therefore requires simulations to include some radiative
transfer. In adition, radiative cooling becomes important in these
regions, as do the potential feedback effects from star-formation
and/or active galactic nuclei. The reliability of the simulations
correspondingly decreases, as the physics behind these processes is
relatively poorly understood, and even harder to model
realistically. However the simulations are not in obvious conflict with
the data, except may be for the very strongest lines, but this is at
least in part a consequence of the relatively poor statistics of these
rarer lines (Gardner et al. 1997; Cen et al. 2003; Nagamine, Springel
\& Hernquist 2004).\\

The metals observed in the quasar spectra were presumably synthesised
in stars. It is possible to constrain the physical properties of the
regions in the IGM that contain metals by comparing the metal-line
column density with the corresponding H{\sc I} column density (Cowie et
al. 1995; Songaila \& Cowie 1996; Dav\'e et al. 1998; Aguirre, Schaye
\& Theuns 2002). The results (Songaila 2001; Schaye et. al. 2003) are
surprising. Metals are detected in gas with density below the cosmic
{\em mean} density, albeit with low but uncertain metallicity $Z\ltsima
10^{-3}Z_\odot$, uncertain filling factor, and with little if any
evidence for evolution with redshift over the observable range of
redshifts $2\le z\le 5$. The low density to which they are detected
suggests that metals were able to escape from galaxies due to some type
of (feedback?)  mechanism. However, the lack of strong evolution in the
metallicity, combined with evidence for vigorous star formation and
hence also metal production over the same redshift interval, suggests
that many metals that are present in the IGM are not seen in the QSO
spectra. In contrast, Adelberger and collaborators (e.g. this volume)
argue that the intimate connection between galaxies and metal lines
seen in their data implies that by and large the observed galaxies are
responsible for the observed metals. The observational situations is
thus a bit open to interpretation, but anyhow the present series of
simulations is not very successful in reproducing the observed
metal-line properties in detail (Aguirre et al. 2005 and this
volume).\\

Observed quasar spectra and mock spectra obtained from numerical
simulations are practically indistinguishable for the untrained eye, at
least for the Hydrogen lines. This makes it possible to use mock
spectra to estimate errors, investigate observational biases, and
constrain and interpret correlations. In addition they make it
possible to examine the level to which the parameters that determine
the IGM properties can be constrained, as well as investigate possible
degeneracies between parameters given the statistics of an actual
observed sample of quasars.\\

Given this, the comparison of observed quasar spectra with mock spectra
obtained from numerical simulations has become a very valuable tool in
the interpretation of absorption line data. Unfortunately the comparison
is often hampered by the different languages employed by observers and
simulators. Simulators are often unaware of the uncertainties due to
continuum-fitting for example, partly because this is often done \lq by
eye\rq, and hence almost impossible to quantify. Few non-simulators
would suspect that \lq adiabatic simulations\rq\, are in fact not
adiabatic at all, what is meant is that there is no radiative
cooling. The aim of this paper is to discuss the ingredients of current
numerical simulations, the differences between popular codes, their
strengths and weaknesses, so that the non-expert can judge which results
are more reliable, and which may be less robust. A more in depth recent
review of numerical simulations is by Bertschinger (1998).\\

\section{Hydrodynamic description}
The aim of a numerical simulation is to follow how initially small
density perturbations grow when they enter into the non-linear
regime. This involves a choice of the background cosmological model in
which the fluctuations grow, for example by specifying the density of
the various constituents in units of the critical density. Next we need
to specify the statistics of the perturbations. In addition to these
physical parameters, numerical limitations also require us to decide
values for some purely numerical parameters, for example the size of
the simulation box and the gravitational softening length. And because
the simulations will only resolve a specific range of scales, we may
also need to specify the effect of \lq sub-grid physics\rq, for example
star formation, that happens below the resolution limit and hence is
not computed but rather is imposed. Other physical processes may have
to be added by hand as well, for example simulations of the IGM often
impose the assumed evolution of the ionising background as a function
of redshift. I will discuss these various parameters that need
specifying, and the corresponding equations in turn, starting with the
background cosmological model. Many more are in for example Peebles
(1993).

\subsection{Background cosmological model}
The evolution of the scale factor of the Universe, $a(t)$, is obtained
from the Friedmann equation,

\begin{equation}
H(t) \equiv {\dot a\over a}\,\,\,\,\,\,\,\hbox{and}\,\,\,\,\,\,\, H^2  = H_0^2\,(\Omega_m/a^3+\Omega_r/a^4+\Omega_k/a^2+\Omega_\Lambda\,)\,,
\end{equation}

where $H(t)$ is the Hubble constant at time $t$, $H_0$ is the Hubble
constant now (when $a=1$), and the various $\Omega$s denote the density
of respectively matter, radiation, curvature, and the vacuum-energy
(cosmological constant) in units of the critical density. For example

\begin{equation}
\Omega_m \equiv {\langle \rho_m\rangle \over \rho_c}\,\,\,\,\,\,\,\hbox{
and}\,\,\,\,\,\,\, \rho_c   = {3\,H_0^2\over 8\pi\,G}\,.
\end{equation}

The scale factor is used to convert physical quantities into co-moving
variables, for example for the position, ${\bf r}$, and velocity, ${\bf
v}$,

\begin{equation}
{\bf r} = a\,{\bf x}\,\,\,\,\,\,\,\hbox{and}\,\,\,\,\,\,\, 
\dot {\bf r} = {\bf v} = \dot a\,{\bf x} + a\,\dot{\bf x}\,.
\end{equation}

The physical velocity ${\bf v}$ is the sum of the Hubble expansion,
$\dot a\,{\bf x}$, and a \lq peculiar velocity\rq, ${\bf v}_p\equiv
a\,\dot{\bf x}$. Most simulations are performed using co-moving
variables such as ${\bf x}$ and $\dot{\bf x}$, but there is wide and
therefore confusing choice of how to convert other physical quantities
into co-moving ones.

\subsection{Equations for the growth of perturbations}
The equations we want to solve are those expressing conservation of
mass, momentum, and energy in a self-gravitating fluid (e.g. Landau \&
Lifschitz 1959).  These are the continuity, Euler and energy equations,
supplemented by Poisson's equation,

\begin{eqnarray}
{\partial\rho\over\partial t} + \nabla(\rho\bfv) &=&0\\
{\partial\over\partial t}\bfv + (\bfv\cdot\nabla)
\bfv&=&-{1\over\rho}\nabla p-\nabla\Phi\label{eq:euler}\\
\rho{\partial\over\partial t} u + \rho\,(\bfv\cdot\nabla) u&=& -p\nabla\bfv
\label{Eq:E}\\
\nabla^2\Phi &=&4\pi G\rho_T\,.
\label{eq:CEP}
\end{eqnarray}

Here, $u$ is the energy per unit mass, $u=p/(\gamma-1)\rho=\boltz
T/(\gamma-1)\mu\mhydro$ of the fluid with density $\rho$, pressure $p$,
temperature $T$, and mean molecular weight $\mu\,\mhydro$. Dark
matter, stars and baryons contribute to the density in the right hand
side of the Poisson equation, but only baryons contribute to the
pressure gradient.

The derivatives with respect to time are assumed to be evaluated at
constant ${\bf r}$ in these Eulerian equations. We can re-write them in
terms of derivatives at constant co-moving position, ${\bf x}$, using
the relation ${\bf r}=a\,{\bf x}$ (see e.g. Peebles 1993 for details)

\begin{eqnarray}
\dot\delta +\nabla\left[(1+\delta)\dot\bfx\right] &=& 0\nn\\
\ddot\bfx + 2H\dot\bfx + (\dot\bfx\cdot\nabla)\dot\bfx
&=&-{1\over a^2}\nabla\Psi - {1\over a^2}\,{\nabla\,p\over\rho}\nn\\
\dot u+ 3H{p\over\rho} + (\dot\bfx\cdot\nabla) u &=& -{p\over\rho} \nabla\dot\bfx\nn\\
\nabla^2\Psi &=&4\pi G\rho_b\delta a^2\,.
\label{eq:nsf}
\end{eqnarray}

The density is written as $\rho(\bfx, t)\equiv
\rho_b(t)\,(1+\delta(\bfx, t))$, where $\rho_b(t)$ is the
time-dependent background density, $\delta \ge -1$ the deviation from
uniformity, and the derivative is
$\nabla\equiv\partial/\partial\bfx$. These equations are valid in a
reference frame at rest with respect to the matter in the Universe,
since the cosmological principle of a homogeneous and isotropic
Universe only applies to such \lq Fundamental Observers\rq\,.\\

If it were not for the explicit time-dependencies arising from the
scale factor, $a(t)$, and background density, $\rho_b(t)$, then the
dynamics expressed by these equations is purely Newtonian. How good
this approximation is, can be estimated by calculating post-Newtonian
corrections arising from General Relativity. These corrections depend
on the size of the region in units of the horizon size squared,
$(l/ct_0)^2$, times the amplitude of the density perturbation on the
scale of the region (see Matarrese \& Terrenova 1996; Takada \&
Futamase 1999; Ellis \& Tsagas 2002 for recent discussions on this, and
Peebles 1993 for more background). In practise this means that the
Newtonian approximation is very good, since in most simulations of the
currently popular models the typical density fluctuations are indeed
small on the scale of the horizon.\\

These differential equations of course implicitly assume that the fluid
variables are everywhere continuous, but that need not be the case. The
conservation equations also allow discontinuous changes, which
correspond to for example {\em contact discontinuities}, through which
the pressure is continuous but the density and thermal energy are not,
and {\em shocks} through which also the pressure is discontinuous (see
e.g. Harlow \& Masden 1971 for an excellent introduction). These more
general solutions are called the Rankine-Hugoniot shock conditions, and
can be solved exactly for one-dimensional discontinuities. The widths
of such discontinuities are of order of the mean-free path of the
particles, which is usually much shorter than what can be resolved in
the simulations and therefore usually neglected. Their widths in the
numerical calculation is then determined by the numerical scheme
employed.\\

It is possible to convert these differential equations into difference
equations, expressing conservation of mass, momentum and energy as the
fluid flows between computational cells. These are {\em Eulerian}
simulations if the cells themselves do not move. In contrast for {\em
Lagrangian} simulations the change in fluid properties is expressed
with respect to cells that move with the fluid. This can be done by
introducing the Lagragian time-derivative, $d/dt\equiv
(\partial/\partial t+\dot{\bf x}\cdot\nabla)$, in terms of which the
continuity changes to

\begin{equation}
{d (1+\delta)\over dt} + (1+\delta)\,\nabla\cdot\dot{\bf x}=0\,.
\end{equation}
A lock keeper who computes the rate of flow of the river past his lock
performs Eulerian calculations, whereas the rafter who notices changes
in the river as the raft moves along with the flow is a Lagrangian
calculator. It is of course also possible to study the flow with
respect to coordinates that move, but not with the fluid velocity:
these are moving-mesh calculations. The numerical implementation of
these equations, and the advantages of the different schemes, are
discussed further below.\\

Equation~(\ref{Eq:E}) expresses entropy conservation, since
$p/\rho^{\gamma}$ is constant along the fluid's
trajectory. Simulations that integrate these equations are therefore
often referred to as {\em adiabatic} simulations but this is a
misnomer since entropy is generated when shocks occur: non-radiative
simulations would be a better name.  The entropy can also change
because of radiative processes which will be discussed below.

\subsection{Initial conditions}
It is usually assumed that density perturbations on a given scale at
sufficiently early times are small. This is true for cold dark matter
models, but not for some more exotic cosmologies. The statistical
properties of the initial conditions (ICs) can be specified in terms
of the Fourier coefficients of for example the density perturbations
$\delta=\rho/\langle\rho\rangle-1$, as,

\begin{equation}
\delta(\bfx,t)=\sum\,\hat\delta(\bfk,t)\exp(i\bfk\cdot\bfx)\,.
\end{equation}

The definition of a {\em Gaussian} random field, is one where the real
and imaginary parts of $\delta(\bfk,t)$ are independently Gaussian
distributed with zero mean and dispersion $P(k)/2$, where $P$ is the
power spectrum of fluctuations (see e.g. Bardeen et al. 1986 for a
comprehensive overview). The power spectrum, $P(k,t)$, only depends on
the magnitude, $k=|\bfk|$, of the wave-vector to guarantee
isotropy. The probability distribution, ${\cal P}$, for the real part
and imaginary parts, $\delta_R$ respectively $\delta_I$, of a Gaussian
field are thus by definition

\begin{equation}
{\cal P} (\delta_R,\delta_I)\,d\delta_R d\delta_I= {1\over
\pi\,P}\,\exp(-{\delta_R^2+\delta_I^2\over P})\,.
\end{equation}

Gaussian statistics may result from the central limit theorem or
inflation (Guth 1981), and observations of the cosmic microwave
background are consistent with Gaussian statistics (e.g. Komatsu et
al. 2003). In the linear regime when the fluctuations are small, the
different $k$-modes grow independently and the linear power-spectrum
can be written as a product of the primordial power-spectrum,
$P(k,t_i)$, and a transfer function, $T(k,t)$,

\begin{equation}
P(k,t)=P(k,t_i)\,T^2(k,t)\,.
\end{equation}

The growth of a perturbation from the primordial epoch at $t=t_i$ to
the early time $t$ depends on the the matter content of the model and a
variety of fitting functions to $T(k,t)$ are in common use. However
when generating IC for a simulation it is equally possible to compute
$T(k,t)$ directly with a Boltzmann code such as for example \cmbfast\,
(Seljak \& Zaldarriaga 1996), and use a tabulated version. Given the
transfer function, the linear power-spectrum is now completely
specified by its primordial shape, $P(k,t_i)$. Symmetry considerations
(Harrison 1970; Zel'dovich 1972) and cosmic inflation (Guth 1981) both
suggest a (nearly) scale-invariant form $P(k,t_i)\propto k^{n}$,
$n\approx 1$. The amplitude of the power-spectrum, traditionally
characterised by the mass variance on cluster scales ($\sigma_8$), {\em
i.e.} in spheres of radius $8h^{-1}\Mpc$, can not be reliably computed
yet, and is left free to fit the data, for example $\sigma_8=0.9$ for a
cluster normalisation (Eke at al. 1996).\\

Once a power-spectrum and its normalisation are chosen, numerical
simulations are set-up by generating a realisation of such a random
field using a pseudo-random number generator (e.g. Press et al. 1992)
and using Lagrangian perturbation theory (the Zel'dovich 1970
approximation; Davis et al. 1995). However, the range of waves that can
be represented in a finite calculation is of course finite (see
e.g. Hockney \& Eastwood 1988 for details). How serious a limitation is
that?  Consider a simulation in a periodic box with co-moving size $L$,
and try representing the linear density field on a regular grid with
grid-spacing $\Delta$.  This grid can represent waves with wavelength
$2\Delta\le\lambda\le L$ or wave-vector $\pi/\Delta\ge k\ge
2\pi/L$. The mean density in the simulation box equals the cosmic mean
because of the inability of the grid to represent waves with
$\lambda\ge L$.  Therefore the simulation inevitably suffers from a
lack of {\em cosmic variance} and the statistics of rare objects which
result from large-scale power, such as very massive objects or very
large voids, will not be represented properly by the ICs -- and hence
by the simulation. The only way to decide how severe a limitation this
is, is to perform simulations with a larger box as well and look for
convergence. The simulation also lacks power below the grid spacing. In
CDM simulations transfer of power is mostly from large to small scales,
so this need not be a major problem. However because the first
structures to collapse do so on very small scales, the lack of
small-scale power may still influence the high-density regions and in
particular the slope of the inner density profile of the dark matter
halos (e.g. Binney \& Knebe 2002). Transfer of power could also result
from baryonic physics, for example a generation of early massive stars
potentially disturbs the gas in a large region around them which might
affect subsequent structure formation.\\

The missing large-scale fluctuations due to a finite box-size also
means that the ICs do not represent the initial correlation function,
$\xi(r)\equiv \langle\rho(\bfx)\,\rho(\bfx+\bfr)\rangle$ accurately.
This can be improved by generating the ICs in real space as pointed-out
by Penn (1997), implemented by Bertschinger (2001), and discussed in
more detail by Sirko (2005)\\

{\em Zoomed initial conditions} A possible way to break the dilemma
between wanting the simulation box to be at the same time large -- for
better statistics and a proper representation of the power-spectrum --
and small -- for better resolution -- is to perform zoomed simulations,
where the best of both worlds results from using high resolution in a
small fraction of a large simulation volume. Reed et al. (2005) used a
hierarchy of such zoomed simulations to obtain a mass resolution of
better than a solar mass in a comfortably large box of size 497
h$^{-1}$Mpc\,. The dynamic range of the initial conditions required is
so large that it becomes impossible to perform the huge fast Fourier
transform when the ICs are generated in Fourier space. The {\sc
grafic2} ICs generator (Bertschinger 2001) is ideal for this type of
problem, because it never requires a large FFT since the ICs are
generated in real space.

\subsection{More baryonic physics}
The equations for the growth of structure given earlier are not enough
for an accurate description of the behaviour of baryons. We need to
add terms that describe the radiative cooling and heating of the gas,
and describe the effect of feedback.\\

{\em Radiative cooling} A collision between an electron and an atom or
ion may be able to excite or ionise the atom if the collision energy is
comparable to, or higher than, the ionisation energy, $E_i$. This
represents a loss term for the thermal energy of the system if the
photon which results from de-excitation or recombination is able to
escape; this is radiative cooling. The cooling rate is proportional to
the density squared since the process involves the collision between
two particles. The temperature dependence can be understood from the
requirement that the collision energy needs to be of order of the
binding energy. Consequently, cooling will become inefficient once the
typical collision energy is far below the ionisation energy,
$\boltz\,T\ll E_i$, in practise $T\le 10^4$~K for cooling by atomic
hydrogen. The gas will become mostly ionised once $\boltz\,T\ge E_i$,
also quenching cooling. Therefore the cooling rate for a given ion will
peak at energies around but $\le E_i$. The net radiative cooling rate
is then the sum of the cooling rates due to all ions combined. The
cooling rate of gas sufficiently enriched by metals will be dominated
by these heavier elements. Dalgarno \& McCray (1972) review these
processes and give fitting formulae for the net cooling rate summed
over all ions assuming ionisation equilibrium.\\

Electrons will loose energy through Coulomb interactions with ions
which is the thermal Bremsstrahlung that makes clusters of galaxies
glow in X-rays. This cooling channel becomes important for $T\ge
10^{6.5}$K. Finally electrons loose energy by inverse
Compton-scattering with photons from the Microwave Background. This
rate depends on the energy density of the CMB and hence increases
rapidly with redshift, $\propto(1+z)^4$.\\

{\em Photoionization} A sufficiently energetic photon of energy
$h\,\nu\ge E_i$ may ionise an atom. The excess energy $h\,\nu-E_i$ goes
into the kinetic energy of the electron representing a heating term;
this is photoionization heating. The rate of photoionizations depends
on the flux of ionising photons, $J(\nu)$ (usually expressed in the
unwieldy units of erg cm$^{-2}$ s$^{-1}$ Hz$^{-1}$ sr$^{-1}$) as

\begin{equation}
\Gamma = \int_{\nu_i}^\infty {4\pi J(\nu)\over
h\nu}\,\sigma_i(\nu)\,d\nu\,,
\end{equation}

where $\sigma$ is the photoionization cross-section, and $E_i=h\nu_i$. 
The change in ionisation balance in a pure hydrogen gas, resulting from
photoionizations and recombinations is given by

\begin{equation}
{d {\rm H{\sc I}}\over dt} = \alpha_{\rm H{\sc I}}\,n_e\,n_{\rm H{\sc
I}}-{\rm H{\sc I}}\,\Gamma
\label{eq:eb}
\end{equation}

where ${\rm HI}\equiv n_{\rm H{\sc I}}/n_{\rm H}$, $n_e$ is the electron number density,
$\alpha$ is the recombination coefficient and I neglected collisional
processes. The corresponding heating rate is

\begin{equation}
\rho\,{du\over dt} = \epsilon\,{\rm H{\sc I}}\,,\,\,\,\,\,\,\,\hbox{where}\,\,\,\,\,\,\, \epsilon = \int_{\nu_i}^\infty {4\pi J(\nu)\over h\nu}\,\sigma(\nu)\,(h\nu-h\nu_i)\,d\nu\,,
\end{equation}

with $\epsilon$ the mean energy that goes into the gas
per photoionzation.\\

When the gas is highly ionised and in photoionization equilibrium
(${d{\rm H{\sc I}}/ dt}=0$), Eq.~(\ref{eq:eb}) simplifies to
$\alpha\,n_{\rm H}^2= \Gamma\,n_{{\rm H{\sc I}}}$, and the heating rate
becomes

\begin{equation}
{du\over dt}\propto {\epsilon\over\Gamma}\,{\alpha}\,\rho\,.
\end{equation}

This shows that the heating rate is {\em independent} of the amplitude
of the ionising background, $J$, (since both $\epsilon$ and $\Gamma$
are $\propto J$) but does depend on the {\em physical} density of the
gas.\\

Other heating mechanisms may be important as well, for example
photoelectric heating by dust grains (Nath, Sethi \& Shchekinov '99;
Inoue \& Kamaya 2003), by X-rays either post (Madau \& Efstathiou '99)
or pre-reionisation (Ricotti \& Ostriker 2004), or cosmic rays (Samui,
Subramanian \& Srianand 2005).\\

The radiative cooling rate depends on the ionisation state of the gas,
since only neutral ions (H{\sc I}, He{\sc I}, He{\sc II}) can cool. The
presence of an ionising background may change the ionisation state
thereby quenching cooling, see {\em e.g.} Efstathiou (1992) for the
effect this has on the formation of dwarf galaxies. This is of course
true for cooling by metal lines as well. Note also that the gas is not
necessarily in ionisation equilibrium: non-equilibrium effects may be
important when the UV-background changes rapidly, for example during
reionisation (Abel \& Haehnelt 2000). All this makes it hard to include
radiative cooling accurately in simulations, with many simulators
preferring to use the equilibrium Dalgarno \& McCray tables which do
not include photoionization.\\

When the photoionization equations are integrated to determine the
ionisation states, one needs to know the temperature dependence of the
atomic parameters, such as recombination and photoionization rates,
over a large range in $T$, given the large temperature range, $10^4{\rm
K}\le T\le 10^8{\rm K}$ encountered in typical simulations. As a
result, there is a variety of fitting formulae for the temperature
dependence of these parameters in common use. A cursory comparison
reveals differences $\ge 60$\%. To get around this simulations could
use tabulated values for these constants, obtained from experimental
measurements or quantum mechanical calculations. The {\sc hazy} website
(http://www.nublado.org/; Ferland 2000) has links to such data
bases. Although this \lq uncertainty\rq\, in the Hydrogen atomic
parameters are due to poor fits rather then poorly known physics, the
atomic parameters for some transitions in metals may be genuinely
uncertain by a sometimes surprisingly large factor.\\

The temperature of the low-density IGM is set by photoionization, if
other heating mechanisms are neglected. Both galaxies and quasars
contribute to $J$ (e.g. Haardt \& Madau 1996), but the relative
contribution is uncertain, as is the spectral shape $J(\nu)$ and
amplitude $J_{21}$ at the Hydrogen ionisation edge (see e.g. Bolton et
al., this volume). Photons emitted by hot shocked gas may also
contribute to the photo-heating rate $\epsilon$ (Miniati et
al. 2004). Most cosmological simulations are too small to have a
representative number of sources of each type from which to compute
$J$, and hence an ionising background is imposed by hand, by specifying
the redshift evolution of $\Gamma$ and $\epsilon$.\\

When the Universe is highly ionised, photoionizations are dominated by
photons with $\nu\gtsima \nu_i$ since the ionisation cross-section,
$\sigma(\nu)$, peaks there. Consequently the heating rate is low. In
contrast during reionisation most photons will cause an ionisation, the
mean energy injection per ionisation increases, and the Universe gets
heated uniformly if one neglects radiative transfer effects
(Miralda-Escud\'e \& Rees 1994). Note that the heating rate depends on
the spectrum of ionising photons, which will change as the source
spectrum is processed by the intervening absorbers.\\

The epoch of reionisation is relatively uncertain, as is the nature of
the sources responsible. If reionisation happens relatively quickly,
then the Universe will become nearly isothermal at that epoch. It is
then a combination of adiabatic cooling and photoionzation-heating that
determines the further thermal evolution, at least in the low-density
IGM ($\rho\le\langle\rho\rangle$ say) where radiative cooling is
unimportant. This establishes a tight temperature-density relation,
$\rho\propto T^{\gamma-1}$ (Hui \& Gnedin 1997; Theuns et al. 1998),
and the slope steepens from isothermal $\gamma=1$ at reionisation to an
asymptotic value $\approx 1/1.7$ as the IGM cools adiabatically. The
steepening results from the fact that the lower-density gas cools more,
and is heated less, since $du/dt\propto \rho$. If this were the whole
story, then the IGM temperature-density relation can be used to
constrain the reionisation epoch (Miralda-Escud\'e \& Rees 1994;
Haehnelt \& Steinmetz 1998; Theuns et. al 2002a).\\

Some regions of the Universe may have been reionised before others if
reionisation is patchy. This will lead to large variations in
temperature during reionisation, and the long thermal time-scales in
the IGM guarantee that some of these temperature fluctuations may still
be observable at later times. Such temperature fluctuations have not
been found yet (Theuns et al. 2002b; Zaldarriaga 2002). The mean free
path of X-ray photons is much larger and one would not expect to see
temperature fluctuations if X-ray heating dominated (Ricotti \&
Ostriker 2004). The thermal state of the IGM could thus plausibly
elucidate the nature of the sources responsible for reionisation.\\

{\em Feedback and subgrid physics} Cosmological hydrodynamical
simulations are far from resolving Giant Molecular Clouds which are the
sites of star formation. In addition they lack important physics for
example dust cooling, or the dynamical effects of magnetic fields and
cosmic rays, {\em etc.}, which are known to be important in the
interstellar medium of the Milky Way, say. Therefore since they cannot
{\em simulate} how cold gas gets converted into stars they use recipes
of how this happens instead, usually of the form where \lq cold dense
gas\rq\, gets converted into stars at a given rate, ${\cal R}$, with a
given efficiency, $\epsilon$, with ${\cal R}$ and $\epsilon$ chosen
after some experimentation. More sophisticated approaches model the
multi-phase interstellar medium as discussed by McKee \& Ostriker
(1997) and more recently by Monaco (2004), by integrating rate
equations that govern the mass exchange between a hot phase (the one
simulated by the hydrodynamical modules of the code), a cold phase, and
stars. This subgrid step is important because it is thought that the
feedback from star formation regulates the conversion from gas into
stars. How this works in practise is not well understood, and
consequently there is a wide variety of implementations of how the
energy released by massive stars and Super Novae couples back to the
gas. Other feedback processes may be important as well, for example
due to AGN.\\

The gas cooling rate and hence the star formation rate, can be
dominated by the metals produced by stars and SNe. But how well are the
metals mixed? And how does this happen? After all, the metallicity range
between the highly super solar SNe ejecta, the interstellar,
inter-cluster and finally intergalactic gas is very large.  In
simulations, the extent of the mixing is mostly determined by the
details of the numerical implementation. It might be possible to get a
handle on this by comparing mock metal spectra to observations, for
example using observations of parallel sight lines in lensed quasars
(Rauch et al. 1999), but the issue can also be addressed in galaxies
(Brighenti \& Mathews 200) and clusters of galaxies (Morris \& Fabian
2003).\\

\section{Numerical simulation codes}
The development of a mature simulation code is usually measured in
person-years. Fortunately several groups have started making their
codes publically available and this, together with the rapid increase
in computer power, has given a huge boost to the simulation industry.

\subsection{Lagrangian versus Eulerian}
An important distinguishing feature between codes is whether they are
Eulerian or Lagrangian, {\em i.e.} whether they follow the fluid with
respect to fixed cells, or whether they use \lq cells\rq\, (particles)
that move with the fluid.\\

{\bf Eulerian codes} can accurately incorporate the Rankine-Hugoniot
shock conditions by treating the boundary between cells as the \lq
membrane\rq\, in a shock tube experiment (the Riemann problem). They
are therefore very good at computing for example the intricate
Kelvin-Helmholtz and Rayleigh-Taylor instabilities that develop when
hot tenuous bubbles expand into a denser atmosphere, and the home pages
of these codes usually contain beautiful examples of such
instabilities. In addition, it is straightforward to distribute the
computational cells over many compute nodes of a parallel computer and
still achieve excellent load balance with a minimum of communication
overhead. The simulations by Cen and Ostriker (1994) use such a code,
in which the numerical resolution is uniform across the domain. Note
that these codes are not exact Riemann solvers because the exact
solution of the jump conditions can only be obtained in one
dimension.\\

Using a uniform mesh has the great advantage of simplicity, but it does
mean that the code may waste computational resources in un-interesting
regions and at the same time lack resolution where the real action
is. Consider for example a large run of $1024^3$ cells in a 100Mpc box,
then the cell size and hence resolution is 100Mpc/1024$\approx$0.1Mpc,
which gives impressive mass resolution in voids, but is not enough to
resolve a galaxy. {\bf Adaptive Mesh Refinement} (AMR) codes try to
avoid using ever more cells by varying the cell size across the grid,
based on some refinement criterion specified by the user. For example
try to keep the mean mass per cell constant, or require the gradient of
a quantity to be sufficiently small across a single cell.  These codes
are significantly more complicated than single grid codes, and need to
take good care of what happens at the interface between smaller and
larger cells. The gains of using fewer cells may be undone by the
overhead associated with the adaptive part of the code, depending on
the problem and the type of refinement criterion employed.  {\sc flash}
(http://flash.uchicago.edu/website/home/; Fryxell et al. 2000) and
\enzo\,(http://cosmos.ucsd.edu/enzo/; Bryan et al. 1995) are publically
available AMR codes, while other AMR codes include {\sc art} (Kravtsov
et al. 2002), {\sc masclet} (Quilis 2004) and {\sc ramses} (Teyssier
2002).\\

Eulerian codes compute how material moves across the grid, and this
can be difficult in cosmology because the high velocity of the cold
material in the IGM means potentially very high Mach number flows.
This problem does not arise in {\bf Lagrangian codes} which simply
follow the fluid as it moves, and this is obviously quite an advantage
for cosmological simulations for the growth of structure. {\em Smoothed
Particle Hydrodynamics} (SPH; see Monaghan 2002 for a review) uses
interpolation on irregularly gridded data to obtain smooth estimates of
hydrodynamical quantities. For example the density, $\rho$, at position
$\bfr_i$ is

\begin{equation}
\rho(\bfr_i) = \sum_j m_j\,W(|\bfr_i-\bfr_j|/h_{ij})\,,
\label{eq:SPH}
\end{equation}

where the sum extends over \lq nearby\rq\, particles
($|\bfr_i-\bfr_j|\le h_{ij}$), $m_j$ is the mass associated with
particle $j$, and $W$ is a smoothly peaked interpolation function. The
spatial resolution is set by $h_i$, which typically is chosen such
that about 40 particles contribute, and the simulations use
linked-lists (Hockney \& Eastwood 1988) or hierarchical trees (Barnes
\& Hut 1986) to identify neighbours. In general this means that the
resolution is higher in high-density regions, and so Lagrangian codes
generally resolve high density regions better than Eulerian codes,
whereas the latter may do better in under-densities. There is a variety
of possibilities for how to convert the hydrodynamical equations into
the type of interpolation in Eq.~(\ref{eq:SPH}).\\

Smooth interpolation cannot handle the discontinuous changes associated
with shocks hence requires the addition of extra terms that act like a
viscosity. Their net effect is to convert ram-pressure into thermal
pressure when the fluid passes through a shock. Such \lq artificial
viscosity\rq\, terms manage to capture shocks within a few resolution
lengths $h$, even for very strong shocks. They increase the pressure
gradient between two rapidly approaching particles to mimic
ram-pressure. The switch that triggers the artificial viscosity can
mistake a shear flow such as occurs in a differentially rotating galaxy
disk for a shock. This has the unwanted side effect of transferring
angular momentum through the disk, but the severity of the effect can
be controlled by increasing the resolution. There is a large number of
SPH codes in use, \gadgetII\, (Springel 2005) can be downloaded from
http://www.mpa-garching.mpg.de/gadget/.\\

Which code/numerical scheme is best really depends largely on the
problem being studied. Zoomed simulations -- in which one is interested
in high resolution in a small fraction of the total simulation volume,
such as simulations of the formation of individual galaxies say,
obviously benefit from the advantages of AMR or SPH simulations over
uniform grid codes. But the shear speed and efficiency of a single grid
code may yet win the race when uniform resolution is sufficient. AMR
and SPH simulations have another big potential advantage: the
possibility to have a hierarchy of time steps, where some regions of
spaces are updated more frequently than others. A super nova (SN)
exploding in a high density region may cause the time step to plunge
dramatically in a small region of space. Clearly it is a great
advantage if it is possible to decrease the time step in the affected
region, but continue striding with large steps outside it. \enzo\, and
most SPH codes use such a hierarchy of time steps, and it leads to a
very substantial time savings. Note that there are numerical issues on
the boundary between the time step zones.\\

The Santa Barbara cluster comparison project (Frenk et al. 1999)
compared the properties of the {\em same} cluster at z=0, simulated
with a variety of codes, and more recently O'Shea et al. (2003)
compared \enzo\, with \gadgetII. Not surprisingly, the more involved
the physics is, the more the results of these codes differ. Whereas the
structure in the collisionless dark matter is very similar in all
codes, there are significant differences in the central gas
distribution because of the different ways that shocks are treated. The
differences would have been much greater had radiative cooling, and/or
stellar feedback been included. \\

\subsection{Performing a simulation}
It is now straightforward to download and compile one or more
simulation codes, and interesting results can be obtained using nothing
more than a standard desktop computer. In contrast, large simulations
may take weeks to run on a large supercomputer producing several Tera
bytes of data in the process. This can make post processing a real
nightmare. Most simulators stick almost religiously to a single code,
because unfortunately there is still a significant overhead in learning
how to use a code. This is largely due to the bewildering variety of
input/output formats and units used. This is exacerbated by the fact
that the initial conditions generator, simulation code, and
visualisation software, usually assume a different set of
units/formats, which is why simulators stick to a given set. The
situation could be improved by a wider use of hierarchical data formats
such as for example HDF5 (http://hdf.ncsa.uiuc.edu/HDF5/), which allow
a much more detailed description of the data than the usual binary
file, without significant overhead.\\

\subsubsection{Parameter choices}
Modern cosmological hydrodynamical simulation codes regularly require
several 10s of lines of input parameters, most of which do not get
quoted in the resulting paper. Some of these are particular to the
numerical algorithm while others describe parameterisations of physical
processes. One of the latter is for example the evolution of the
ionising background, and it will be clear from the previous discussion
that the subgrid modelling inevitably involves a lot of parameter
choices.  \\

\subsubsection{Numerical Issues}
{\em Small scale structure and gas cooling} The dark matter power
spectrum may have power on all scales and one should therefore be clear
what is meant when one claims \lq convergence\rq,\, of a
simulation. This problem becomes more accurate when baryons are
included. Gas cooling is strongly density dependent, hence the fraction
of gas that is able to cool depends on resolution. With increasing
resolution, a numerical simulation will predict that a larger fraction
of gas will cool. This is in fact the right answer, and it has long
been known that without feedback the Universe suffers from a cooling
catastrophe where most baryons cool, clearly in contradiction with
observations (e.g. Balogh et al. 2001 for a recent discussion).
However it does mean that simulations that do not include some type of
feedback simulate a physical problem that {\em does not have a
solution}, with the amount of gas cooling determined by
resolution. Claiming that these simulations are resolved is wrong. \\

{\em Over cooling at interface between dense and tenuous gas} The
density in SPH simulations is smoothly varying.  Therefore the presence
of a dense cold cloud may lead to an over estimate of the density of
surrounding tenuous gas, which in turn will lead to artificially
increased cooling. This is a real problem in SPH simulations. A
possible fix is to decouple hot and cold gas when computing SPH
estimates of quantities, but this is at the expense of altering the
equations in an ad hoc way. Grid-based codes suffer to a lesser extent,
but the boundary cells between the two phases may still overestimate
$\rho$ and hence the cooling rate.\\

{\em Time stepping} The time-scale of some physical processes, such as
for example the cooling time, or the lifetime of massive stars, may be
so short that they are not resolved in a cosmological simulation.  This
is usually circumvented by integrating the equations using an implicit
solver, which at least guarantees stability. However the net outcome of
the intricate feedback loop of gas cooling leading to star formation
and SNe explosions again heating the gas, may give significantly
different answers depending on the time step. For example suppose that
during a long time step a lot of gas cools, leading to a large burst of
star formation at the end. With a much smaller time step, feedback from
star formation could have heated the gas and prevented more gas from
cooling, implying a much smaller star formation efficiency. Models that
treat this cycle partly as sub-grid physics can get around this by
taking small time steps in the sub-grid physics, but larger ones in the
more expensive hydrodynamical part of the calculation. Again the
downside is that the equations are not actually simulated by the
hydrodynamical code anymore but rather \lq solved\rq\, by hand.  \\

{\em Feedback} The net effect of a SN explosion depends strongly on the
density of the gas it goes off in. In a high-density region most of the
SN energy may be lost to radiation, whereas the material in a lower
density region may be swept-up and driven out of the potential well of
the galaxy. The real ISM of a galaxy is a much more dynamic and complex
multi-phase medium with both dense and tenuous phases, and cannot be
resolved yet in cosmological simulations. Note also that cosmic rays,
not usually included in simulations, are thought to provide a
significant fraction of the pressure support, at least in the Milky
Way. Given these limitations, it is not clear how we can realistically
expect that the SN explosions in a simulated galaxy power the feedback
cycle correctly. Again it might be necessary to relegate most of this
physics to the subgrid domain, where we simply impose an answer instead
of trying to obtain it by integrating the proper hydrodynamical
equations (Yepes et al. 1997).\\

{\em Metal enrichment and cooling} Simulations that include metal
production usually add these metals to cells or particles near the
site of star formation. Lack of resolution means that the mixing from
high density SN shells to the much lower ISM densities is a result of
the numerical implementation of the mixing, it is not {\em
computed}. The metals are associated with a given particle in SPH
simulations, which is in keeping with its Lagrangian nature. If this
metal enriched particle is swept out of the galaxy by a galactic wind,
it may find itself surrounded by zero metallicity gas falling into the
galaxy for the first time. The {\em mean} metallicity at this radius
may be very low, but it is the result of many particles having zero
metallicity, and few having high metallicity. This may already seem
awkward, but a problem arises if we want to compute the cooling rate
due to metals. It may be low in the zero metallicity gas, but high for
the single particle. Yet this is mostly a result of the way we mixed,
or rather did not mix metals, it was not the result of a
computation. Suppose that as a result of metal cooling the expanding
shell fragments: should we trust this outcome?  In an Eulerian code,
metals would mix when the enriched gas rises out of the galaxy, and
this may give significantly different answers.\\

{\em Subgrid physics} An increasingly large fraction of the numerical
calculations performed during a simulation is treated as subgrid
physics, due to limitations of missing physics and lack of resolution
in the hydrodynamical modules of the code. This increasingly blurs the
distinction between \lq semi-analytical calculations\rq\, and numerical
simulations. It also means that it becomes difficult to quantify the
robustness of results obtained from simulations, because some results
may result directly from the subgrid physics module, and not from the
integration of hydrodynamical equations. 

\section{Comparing mock spectra to data}
To obtain a high-resolution, high signal-to-noise quasar spectrum
requires many hours on a large telescope and is therefore an expensive
commodity. Consequently, observers will spend a significant amount of
time reducing the data to obtain most out of the spectrum. However, it
is easy to generate many hundred mock spectra from a simulation, and
it is impossible to analyse them in the same detail. This is
unfortunate because we would clearly like to treat observed and mock
spectra in the same way.\\

It is possible to generate mock spectra for comparison with a given
quasar spectra, by creating spectra with the same wavelength extent,
pixel size, and signal-to-noise as function of $\lambda$ and flux. To
do so we compute these spectra as the simulation is running, which
allows us to take the rapid redshift evolution into account
(e.g. Aguirre, Schaye \& Theuns 2002). Note that our mock spectra are
periodic with period equal to the box size. We cycle these short
periodic spectra and stitch them together where there is little
absorption to avoid discontinuities. This inevitably introduces
undesirable periodicities in the mock spectra. Of course this step is
only necessary if the spatial location of the pixels is important.\\

Although this gives observed and mock spectra with very similar data
quality and noise properties, observed spectra also need to be
continuum fitted, to take into account the quasar's continuum
variation, and the wavelength dependence of the sensitivity of the
detector. At present this is one of the most important sources of
uncertainty since this process is often done by eye and hence cannot be
modelled. The properties of the mock spectra also depend on the assumed
amplitude of the relatively poorly known amplitude of the ionising
background. It has become customary to get around this uncertainty by
scaling the mock spectra so that they have the same mean flux, $\langle
F(z)\rangle$. Unfortunately $\langle F(z)\rangle$ is difficult to
measure and different analysis procedures give significantly different
values (e.g. Bernardi et al 2003). Note that at low $z$ a non-negligible
part of the absorption is due to metals, which may not be present in
the mock spectra.\\

Quasar spectra can be analysed by fitting Voigt profiles to the
absorption lines (e.g. using {\sc vpfit},
http://www.ast.cam.ac.uk/$\sim$rfc/vpfit.html). This produces a line list
with three numbers, (1) a central wavelength, (2) width and (3) column
density for each line. It has been shown that the statistical
properties of the lines depend on the physics of the IGM. For example
the fraction of strong lines increases with the amount of large-scale
power, whereas the line-widths depend on the temperature-density
relation. Some lines can be fit well by a small number of Voigt
components, but others differ significantly form the Voigt profile
shape and hence require a large number of components. This
decomposition in Voigt components is not unique, since Gaussians are
not orthogonal basis functions, and will thus depend to some extent on
how lines are fit. Different fitting routines ({\sc vpfit}, {\sc
fitlyman}, {\sc autovp}) will produce different line-lists, definitely
on a line-by-line basis but also statistically. A large fraction of the
fitted lines will occur in blends and their parameters are therefore
not independent. This complicates the statistical comparison between
line-lists from data and mock spectra.\\

The central part of a sufficiently strong line becomes effectively
black, and therefore gives little hint about the absorption profile
there. It is sometimes possible to constrain such lines using
higher-order transitions. However this makes it even harder to compare
line-lists from observed and mock spectra, because the decision to
make use of the higher-order lines for a given absorption line is not
automated hence introduces a further subjective step in the
analysis.\\

A more objective way of analysing the spectra is in terms of pixel
statistics, for example the flux probability distribution (PDF) ${\cal
P}(F)\,dF$, the fraction of pixels with a given flux $F$ (e.g. McDonald
et al. 2000). Continuum fitting and thermal broadening introduce
correlations between the pixels, which need to be taken into
account. For example although the {\em number} of pixels in the
Lyman-$\alpha$ forest part of a spectrum is very large, one should not
be fooled into believing that this implies great statistics for the PDF
since these pixels occur in a relatively small number of {\em
lines}. Analysis of many mock spectra can be of great help to obtain
more realistic error bars, but note that some statistics may be
adversely affected by very strong lines which the simulations may not
model particularly well. Analysis of pixels is very fast, and makes it
possible to look at all transitions from a given redshift, including
higher-order Lyman-$\alpha$ transitions but also metal line
transitions. This is exploited to great effect in the pixel optical
depth method (Aguirre, Schaye \& Theuns 2002 and references therein).\\

Orthogonal to a pixel analysis is to directly compute the
power-spectrum of the flux, $P_F(k)$ (Croft et al. 1998, Mcdonald et
al.  2000). Some parts of the spectrum may need to be excluded because
of the presence of gaps in the data, or the presence of strong metal
absorption lines, but this can be done with the Lomb periodogram (Press
et al. 1992). The quasar sightline samples the 1D power-spectrum, which
is an integral over the 3D power-spectrum and therefore sensitive to
small scale power. One way to see that this matters is to realise that
the total column density of a single strong line can be a significant
fraction of the column density of the whole spectrum.  This illustrates
that the power-spectrum could be sensitive to the number and shape of
strong lines (Choudhury, Srianand \& Padmanabhan 2001; Viel et
al. 2004). Of course the number and properties of these strong lines
will depend on cosmological parameters as well and therefore $P_F(k)$
can still be used to constrain them (Mandelbaum et al. 2003).\\

Another alternative is to decompose the spectrum in true basis
functions, for example using wavelets (Meiksin 2000; Theuns \& Zaroubi
2000). This method is fast and objective, but the interpretation of the
wavelet coefficients may be less intuitive. Pixel correlations again
introduce correlations in wavelet coefficients.\\

It is important to realise that the quasar's flux spectrum depends on
many parameters, some of which may be almost degenerate. For example
there is a near degeneracy between the amplitude of the matter
power-spectrum and the (inverse of the) amplitude of the ionising
background. Conversely changing some cosmological parameters may have
unexpected side effects: for example changing $\Omega_m$ changes the
temperature of the IGM, because the models specify the photo-heating
{\em rate} $\epsilon$, and hence $T$ depends on the age of the
Universe, which depends on $\Omega_m$. It does not stop there because
changing $T$ changes the neutral fraction, hence the flux
power-spectrum, hence the inferred matter power spectrum.

\section{Recent results}
{\em Temperature of the IGM} Several groups obtain temperatures $T\sim
10^4$K at $z=3$ at the mean density, from analysing line shapes (Schaye
et al 1999, 2000; Ricotti, Gnedin \& Shull 2000; Bryan \& Machacek
2000; McDonald et al. 2001). The underlying physics uses the fact that
there is a minimum line width at a given column density due to thermal
smoothing. Of course the occasional \lq erroneous\rq\, lines introduced
by the fitting algorithm may yet violate this limit, but one does
expect a rather well defined lower limit to the line width, $b$, as
function of column density, $N_{\rm H}$. Such a cut-off had in fact
already been noticed in the observed $b-N_{\rm H}$ scatter plots (Lu et
al. 1996). The groups make different predictions for the evolution
$T(z)$ but these need not be inconsistent since different methods will
measure and weigh $T$ at different densities differently.  A
temperature $\sim 10^4$K is expected for photo-heated gas, but only if
reionisation occurred relatively late (Theuns et al 2002a). We will
need to look for alternative heat sources if the early epoch of
reionisation advocated by WMAP (Bennett et al. 2003) holds. There is no
indication for spatial fluctuations in the $T$ distribution which might
be expected from patchy reionisation (Theuns et al. 2002b; Zaldarriaga
2002), but the existing noisy estimates were based on fewer spectra
than are available now. There is an indication for an increase in $T$
around $z=3.2$ which might signal HeII reionisation (Theuns et al
2002d).\\

{\em Amplitude and evolution of the ionising background} The amplitude
$\Gamma$ of the ionising background can be estimated by comparing the
mean flux, $\langle F\rangle$, in simulations to the observed
mean. However, $\langle F\rangle$ is difficult to determine
observationally because of systematic uncertainties in the continuum
fitting. Continua are fitted to high resolution spectra by assuming
that the flux recovers to the continuum level between the lines. This
somewhat subjective procedure becomes increasingly suspect at higher
redshift where simulations show that the flux hardly ever fully
recovers. Lower resolution flux calibrated spectra are continuum fitted
by extrapolating from longer wavelengths. This goes wrong at lower
redshifts where a fraction of the absorption is due to metal lines,
which are not resolved. The differences in the observed evolution of
$\langle F(z)\rangle$ between the two methods is uncomfortably large
(e.g. Bernardi et al. 2003). The determination of $\langle F\rangle$
from simulations is also not straightforward. For a given set of
cosmological parameters, $\langle F\rangle$ will decrease both when
increasing numerical resolution and when increasing the box size at
given resolution. In both cases, more gas collapses to higher densities
which decreases the net absorption. Therefore one needs simulations
with a large dynamic range, since high resolution and large box size
put orthogonal demands on the simulation set-up. Note also that the
value of $\Gamma$ required in a simulation to reproduce a given mean
flux, depends on many other simulations parameters, notably the
temperature $T$ of the gas, and the amplitude, $\sigma_8$, of the
fluctuations. Therefore one should avoid the vicious circle of first
measuring $\sigma_8$ by assuming $\Gamma$, then determining $\Gamma$
using the value $\sigma_8$ found earlier.\\

{\em Metals} The evolution of the metallicity of the IGM as determined
either from lines (Songaila 2001) or pixels (Schaye et al. 2003), is
surprisingly weak, with little if any change between redshifts 2 and
5. Simulations have been instrumental in interpreting the distribution
of metals at lower densities, using pixel statistics. Note that strong
metal lines are relatively rare. This makes them hard to simulate
because it could be that the galaxies that produce them are also rare,
and might be absent in the small simulation boxes that have enough
resolution to simulate the enrichment process. The stronger lines also
tend to occur in rare \lq metal systems\rq\, which the simulations that
include metal enrichment do not reproduce. A major uncertainty in
converting the observed metal transitions into a metallicity is the
largely unknown shape, amplitude, and coherence of the UVB at the
higher energies relevant to the metallic ions. It will be interesting
to see whether the new generation of simulations is able to distribute
metals produced in forming galaxies, in a way consistent with the data
(Theuns et al. 2002c; Springel \& Hernquist 2003, Aguirre et al. 2005).\\

{\em Power spectrum} The properties of the low-density IGM depends on
small-scale structures when the fluctuations are still small and is
therefore very well placed to probe the shape and amplitude of the mass
power-spectrum on small scales, and hence address such fundamental
issues such as the index of the primordial power-spectrum, the neutrino
mass and the nature of dark matter including the masses of potential
warm dark matter candidates (e.g. Seljak et al. 2005). It might even be
possible to constrain the geometry of the Universe comparing clustering
along and perpendicular to the line of sight (the Alcock-Pacynski test,
Hui et al. 1999; Rollinde et al. 2003, Lidz et al 2003).\\

{\em In summary} Quasar spectra are a very sensitive probe of the
physical state of the distant intergalactic medium, which we can in
principle use to learn about reionisation, galaxy formation, the
large-scale distribution of matter, the nature of dark matter, and the
geometry of the Universe. It is important to understand to what extent
the Lyman-$\alpha$ forest can live-up to this promise and simulation
are well suited to investigate this.

\begin{acknowledgments}
I thank the organisers for a very stimulating meeting, the IAU for
travel support, PPARC for the award of an Advanced Fellowship, and my
collaborators to allow me to show our results.
\end{acknowledgments}

\end{document}